\documentclass[aps,onecolumn,showpacs]{revtex4}
\usepackage{amsmath,graphicx,subfig}
\usepackage{caption}
\usepackage{parskip}
\usepackage{amssymb,amsbsy}
\begin{document}
\begin{center}{\large{\bf Partially massless theory in three dimensions and self-dual massive gravity}}
\vskip 0.5cm
Daniel Galviz and Adel Khoudeir
\\ \vskip 0.2cm
{\it Centro de F\'{\i}sica Fundamental, Departamento de F\'{\i}sica, \\
Facultad de Ciencias, Universidad de Los Andes, M\'erida 5101, Venezuela }
\vskip 0.3cm
\end{center}

\vskip 1cm

\begin{center}
ABSTRACT
\end{center}
Partially massless theory in three dimensions is revisited and its relationship with the self-dual massive gravity is considered. The only mode of the partially massless theory is shown explicitly through an action for a scalar field on (A)dS background. This fact can be generalized to higher dimensions. This degree of freedom is altered when a triadic Chern–Simons is introduced, giving rise to the self-dual massive gravity on (A)dS background. We present another physical system with partially massless symmetry and its connection with topologically massive gravity is discussed
\vskip 1cm



Published: Modern Physics Letters A Vol. 33, No. 12, 1850067 (Apr 20, 2018).

\section{Introduction}

Partially massless is an interesting property of massive spin-2 on a gravitational background with cosmological constant \cite{Deser-Nepo} and also for higher spin \cite{Deser-Waldron}. It is characterized by the appearance of a novel gauge symmetry when the mass parameter in the Fierz-Pauli action is fixed to certain value of the cosmological constant ($\mu = \frac{2}{D - 1}\Lambda$).This is an alluring fact in the issue of the cosmological constant problem. The gauge parameter is a scalar, which allows the removal of the helicity zero
mode and then reduces by one the number of degrees of freedom of the massive spin-2 ﬁeld. The partially massless theory has aspects very similar toto the Maxwell theory \cite{DW1}, its action can be described in terms of a gauge invariant field strength, conformally invariant \cite{DW2}, stable \cite{DeWa} and enjoying duality symmetry \cite{DW3} \cite{Hint1}. These facts suggest a non linear generalization of the partially massless theory, but only obstructions and no-go theorems which hamper this construction \cite{varios} has been developed. Despite this drawback, there is no conclusive inference until now. Recently, massive graviton with partially massless symmetry beyond the usual (A)dS background have been found \cite{BDH} and a nonlinear bimetric model with partially massless symmetry has been proposed \cite{AH}. On the other hand, in three dimensions, it is possible the formulation of a non linear (quadratic in curvature), covariant version for massive spin-2, known as new massive gravity \cite{BHT}, where its linearization on background (A)dS leads to the partially massless phenomena in three dimensions \cite{BHT2}. However, at a nonlinear level, the partially massless symmetry appears as an unstable point in the non-perturbative canonical structure \cite{Blago&Cvetkovic}, i. e., the partially massless symmetry appears as an accident in the procedure of linearization \cite{HRTZ}. A similar phenomenon appears in the linearization of the lim $\mu \rightarrow \infty$ of the new massive gravity \cite{Deser} (and also in the limit $\mu \rightarrow \infty$ of the so-called topologically new massive gravity \cite{tnmg}), where a conformal symmetry appears, absent at a nonlinear level \cite{Deser-Ertl-Grumiller}. Also, no-go arguments against the construction of a fully nonlinear partially massless theory in three dimensions have been advocated \cite{Alexandrov-Deffayet}. Nevertheless, we consider the partially massless phenomenon at a linearized level on (A)dS, in three dimesions and find out some interesting results. An attempt to consider the partially massless theory in three dimensions, in another context, was considered in ref.\cite{GJMS}.

In this paper, we will consider linearized partially massless theory in three dimensions from the point of view of frame-like description and show explicitly that this describes one scalar mode. The partially massless on (A)dS for higher spin symmetric fields using a frame-like formulation, for arbitrary dimensions, has been developed in ref.\cite{SV}. The action associated with this degree of freedom can be seen as the minimal coupling of the (trivial in flat spacetime) linearized three dimensional Einstein-Hilbert action on (A)dS gravitational background. It is known that fields on (A)dS background behave different as they are in flat spacetime (\cite{BMV}) and partially massless theory in three dimensions is an example of this fact. The generalization of this behavior for mixed symmetry fields in higher dimensions is considered. We will see that there exist a relationship with the self-dual massive gravity \cite{AK}. The scalar excitation of the partially massless is transmuted in a massive spin-2 field when it is coupled, in the frame-like description, with a triadic Chern-Simons action, resulting in an equivalent description of the self-dual massive gravity from which we can obtain a second-order formulation with explicit partially massless symmetry. We show that this last formulation is the partially massless version on a (A)dS background of a second-order system formulated in ref.\cite{AAK}. In a flat spacetime, from this second-order formulation, we can reach the topologically massive gravity \cite{DJT}, i. e., the self-dual massive gravity is equivalent(\cite{AK},\cite{ArKhSt}) to the topologically massive gravity. But as the topologically massive gravity on a (A)dS does not possess partially massless symmetry, we will see that its equivalence with the self-dual massive gravity is broken at this level.

\section{Partially Massless in Three Dimensions}

In this section, we consider the partially massless theory in three dimensions from a frame-like perspective instead of a linearized metric formulation.
To start with, we write the first order formalism for massive spin-2 in (A)dS background in three dimensions, where the independent fields are $e_{mn}$ and  $\omega_{mn}$. These entities do not have any symmetry in their indices. We denote the metric of the (A)dS background as $g$ and the covariant derivative on this background by $\nabla$. We use the  convention that $[\nabla_m , \nabla_ n]v_p = R_{mnpq}v^q$ with $R_{mnpq} = \Lambda (g_{mp}g_{nq} - g_{np}g_{mq})$ and a metric with a signature mostly positive. The action is

\begin{equation}
I = \int d^3 x \sqrt{-g} [\omega_{mq} \eta^{mnp}g^{qr}\nabla_n e_{pr} - \frac{1}{2}(\omega_{mn}\omega^{nm} - \omega^2) + \frac{1}{2} \Lambda  (e_{mn}e^{nm} - e^2) - \mu^2 (e_{mn}e^{nm} - e^2) ],
\end{equation}
where $\mu$ is the mass parameter of the massive spin-2 field. The mass term spoils diffeomorphism and local Lorentz symmetries ($\delta e_{mn} = \nabla_m \zeta_n + \epsilon_{mnr}l^r$ and $\delta \omega_{mn} = \nabla_m l_n - \Lambda\epsilon_{mnr}\zeta^r$). The field equations obtained from independent variations on $e_{mn}$ and $\omega_{mn}$ are
\begin{equation}\label{2}
\eta^{mpq}\nabla_p \omega_{q}    ^n + [\Lambda - \mu^2] (e^{nm} - g^{mn}e)  = 0
\end{equation}
and
\begin{equation}\label{3}
\eta^{mpq}\nabla_p e_q ^n - (\omega^{nm} - g^{mn}\omega) = 0,
\end{equation}
where $\eta^{mnp} \equiv \frac{1}{\sqrt{-g}}\epsilon^{mnp}$. This last equation can be solved for $\omega_{mn}$ in terms of the dreibein giving the following expression :
\begin{equation}\label{4}
\omega^{mn} = \eta^{npq}\nabla_p e_q  ^m - \frac{1}{2}g^{mn}\eta^{pqr}\nabla_p e_{qr} \quad \equiv W^{mn}_{(e)}
\end{equation}
and after substituting into the action and decomposing the dreibein in its symmetric and antisymmetric components as $e_{mn} = h_{mn} + \eta_{mnp}v^p$ with $h_{mn} = h_{nm}$, we obtain
\begin{eqnarray}
I = \int d^3 x \sqrt{-g}[ &-&\frac{1}{2}\nabla_p h_{mn} \nabla^p h^{mn} + \frac{1}{2} \nabla_p h \nabla_p h - \nabla_m h \nabla_n h^{mn} + \nabla_n h_{mp} \nabla^p h^{mn} {\nonumber} \\
&+& 2\Lambda h_{mn}(h^{mn} - \frac{1}{2}g^{mn}h) - \frac{1}{2}\mu^2 (h_{mn}h^{mn} - h^2) - \Lambda v_{m}v^{m} ].
\end{eqnarray}

The symmetric and antisymmetric components of the dreibein are decoupled. As is well known, this action lead to the propagation of two local degrees of freedom. In three dimensions, the transverse traceless spatial component of $h_{mn}$ $(h_{ij}^{Tt}$) vanishes identically and these excitations are scalar modes, specifically, they are longitudinal components of $h_{ij}$.

The partially massless condition in three dimensions is specified by the following condition
\begin{equation}
\mu^2 = \Lambda.
\end{equation}
In consequence, the partially massless action in three dimensions is simply described by the action:
\begin{equation}\label{7}
I = \int d^3 x [\omega_{mq} \epsilon^{mnp}\nabla_n e_{p}^q - \frac{1}{2}\sqrt{-g}(\omega_{mn}\omega^{nm} - \omega^2)].
\end{equation}
It looks like the linearized Einstein-Hilbert action coupled minimally on a gravitational background. The field equation which is obtained from this action, after making independent variations on $e_m ^a$ is
\begin{equation}\label{8}
\eta^{mnp}\nabla_n \omega_p ^q  = 0,
\end{equation}
while equation (\ref{3}) (obtained after independent variations on $\omega_m ^a$) is not altered. If we substitute (\ref{4}) into (\ref{7}), we obtain
\begin{eqnarray}
I = \int d^3 x \sqrt{-g}[ &-&\frac{1}{2}\nabla_p h_{mn} \nabla^p h^{mn} + \frac{1}{2} \nabla_p h \nabla_p h - \nabla_m h \nabla_n h^{mn} + \nabla_n h_{mp} \nabla^p h^{mn} {\nonumber} \\
&+& \frac{3}{2}\Lambda h_{mn}h^{mn} - \frac{1}{2}h^2  - \Lambda v_{m}v^{m} ],
\end{eqnarray}
which is just the second-order action for partially massless theory in three dimensions, in terms of the symmetric field $h_{mn}$ (the field $v_{m}$ is decoupled). This action is invariant under
\begin{eqnarray}\label{sym}
\delta e_{mn} &=& \nabla_m \nabla_n \zeta + \Lambda g_{mn} \zeta = \delta h_{mn}; \quad  \delta f_{mn} = 0 {\nonumber} \\
\delta \omega_{mn} &= 0&,
\end{eqnarray}
which characterize the gauge transformations of the partially massless theory in three dimensions.

The solution of equation (\ref{8}) is
\begin{equation}\label{11}
\omega_{mn} = \nabla_m \nabla_n \varphi + \Lambda g_{mn} \varphi ,
\end{equation}
where $\varphi$ is a scalar field. Substitution of this solution in the first-order action of partially massless (\ref{7}) show us the existence of one scalar excitation:
\begin{equation}\label{12}
I = \Lambda \int d^3 x \sqrt{-g} [ - g^{mn}\nabla_m \varphi \nabla_n \varphi + 3 \Lambda \varphi^2 ].
\end{equation}
Also, this scalar excitation can be illustrated by analyzing the field equations of the partially massless theory. The antisymmetric part of the field equation (\ref{8}) tell us that $\nabla_m \omega_n ^n - \nabla^n \omega_{mn} = 0$, while the covariant divergence $\nabla_ m$ of (eq. \ref{8}) leads to the vanishing of the antisymmetric part of $\omega_{mn}$. Now, by taking the covariant divergence to (eq. \ref{3}), we obtain that the antisymmetric component of $e_{mn}$ vanishes on-shell, i. e. $e_{mn} \equiv h_{mn} = h_{nm}$ and considering the antisymmetry of the field equation (\ref{3}), the constraint $\nabla^n h_{mn} - \nabla_m h = 0$ emerges. Finally, the trace of (eq. \ref{3}) conduce to the vanishing of the trace of $\omega_{mn}$, i. e. $\omega_{m}^m = 0$. Since the solution for partially massless theory has been found (eq. (\ref{11})), we have that
\begin{equation}
\omega_m ^m = 0 \Rightarrow (\nabla^2 + 3\Lambda ) \varphi = 0 ,
\end{equation}
which is just the field equation what one derives from action (\ref{12}).

If the spacetime is flat, the solution of (\ref{8}) is a pure gauge ($\omega_{mn} = \partial_m \lambda_n$) which leads to the triviality of the action, i.e. the absence of massless graviton in three dimensions. We have seen that the partially massless action in three dimensions looks as if the trivial massless spin-2 action in three dimensions is turned on when it is minimally coupled on (A)dS gravitational background and illustrates the fact that massless fields on (A)dS spacetime do have different behavior than that shown in flat spacetime \cite{BMV}. This phenomenon also occurs in dimensions greater than three. For instance, in four dimensions, it is well known that the massless Curtright field ($T_{mn}^p$)(\cite{Curtright} does not propagate on flat spacetime and can be described by the following first-order action (\cite{Z&BK}, \cite{BrAK})
\begin{equation}\label{14}
I = \int d^4 x [\omega_m ^r \epsilon^{mnpq}\partial_n T_{pq} ^r - \frac{1}{2}(\omega_{mn}\omega^{nm} - \omega^2)],
\end{equation}
where $\omega_{mn}$ plays the role of a Lagrange multiplier, which can be eliminated using and solving its field equation. After substituting its value into the action, the second-order action for the Curtright field is obtained. Now, we consider the minimal coupling of the action (\ref{14}) on  (A)dS gravitational background:
\begin{equation}\label{15}
I = \int d^4 x [\omega_m ^r \eta^{mnpq}\nabla_n T_{pq} ^r - \frac{1}{2}\sqrt{-g}(g^{mp}g^{nq}\omega_{mn}\omega^{qp} - (g^{mn}\omega_{mn})^2)],
\end{equation}
which has invariance under the following novel gauge transformations:
\begin{eqnarray}
\delta T_{mn}^p &=& g^{pq}(\nabla_m \nabla_q \zeta_n - \nabla_n \nabla_q \zeta_m ) + \frac{\Lambda}{3} (\delta_m ^p \zeta_n - \delta_n ^p \zeta_m) , {\nonumber} \\
\delta \omega_{mn} &=& 0 .
\end{eqnarray}

This symmetry can be considered as a generalization of the partially massless gauge transformation for the Curtright field. It is worth recalling that this symmetry is characterized by a vector parameter instead of a scalar. The field equation obtained by independent variations on $T_{pq}^r$;
\begin{equation}
\eta^{mnpq}\nabla_m \omega_n ^r  = 0,
\end{equation}
has the following solution:
\begin{equation}
\omega_{nr} =  \nabla_n \nabla_r \varphi + \frac{1}{3}\Lambda g_{nr} \varphi
\end{equation}

and after inserting into (\ref{15}), we arrive to
\begin{equation}
I = \Lambda\int d^4 x \sqrt{-g} [-\frac{1}{2}g^{mn}\nabla_m \varphi \nabla_n \varphi + \frac{2}{3}\Lambda \varphi^2].
\end{equation}
The extension to higher dimension with general mixed symmetry fields is given by the following action
\begin{equation}
I = \int d^D x [\omega_{mq} \eta^{mnp_1 ... p_{D-2}}\nabla_n T_{p_1 ...p_{D-2}} ^q - \frac{1}{2}\sqrt{-g}(g^{mp}g^{nq}\omega_{mn}\omega^{qp} - (g^{mn}\omega_{mn})^2)]
\end{equation}
and the solution of $\eta^{mnp_1 ... p_{D-2}}\nabla_m \omega_n ^q  = 0$ is $\omega_{nr} =  \nabla_n \nabla_r \varphi + \frac{2}{(D-2)(D-1)}\Lambda g_{nr} \varphi$, which leads to the action
\begin{equation}
I = \Lambda\int d^D x \sqrt{-g} [-\frac{1}{(D-2)}g^{mn}\nabla_m \varphi \nabla_n \varphi + \frac{2D}{(D-2)^2(D-1)}\Lambda \varphi^2].
\end{equation}
Now, we return to three dimensions. In Sec. 3, the dynamical self-dual massive gravity \cite{AK} on (A)dS is considered and linked with the partially massless theory through a triadic Chern-Simons term.

\section{Self-Dual Massive Gravity on (A)dS Background}

In this section, we will establish the relationship between the self-dual massive gravity and the partially massless theory in three dimensions. We start by revising the self-dual on (A)dS, which has been considered in ref.\cite{arias-gaitan} and \cite{morand}. First of all, we will check on the propagation of only one degree of freedom on (A)dS. We denote the self-dual massive spin-2 by $w_{mn}$ and the first order action for $w_{mn}$ minimally coupled is written down as
\begin{equation}\label{19}
I = \int d^3 x \sqrt{-g}[\frac{1}{2}w_m ^q \eta^{mnp}\nabla_n w_p ^q -\frac{1}{2} \sqrt{-g}\mu g^{mq}g^{np}(w_{mn}w_{pq} - w_{mq}w_{np})].
\end{equation}
The field equation is
\begin{equation}
E^{mq} \equiv \epsilon^{mnp}\nabla_n w_p ^q - \mu\sqrt{-g}(w^{qm} - g^{mq}w) = 0.
\end{equation}
By considering the trace, we obtain
\begin{equation}\label{21}
w = - \frac{1}{2\mu}\eta^{pqr}\nabla_p w_{qr},
\end{equation}
while applying a covariant divergence yields
\begin{equation}
\nabla_m w^{mn} - g^{mn}\nabla_m w = - \frac{\Lambda}{\mu}\eta^{npq}w_{pq}
\end{equation}
and considering the antisymmetric part of the field equation leads to
\begin{equation}
\nabla_m w^{mn} - g^{mn}\nabla_m w = \mu \eta^{npq}w_{pq}.
\end{equation}
 The last two equations show
 \begin{equation}
 ( \mu^2 + \Lambda )\eta^{npq}w_{pq} = 0.
 \end{equation}
For dS spacetime $\Lambda > 0$ and avoiding the Breitenlohner-Freedman bound: $\mu^2 = - \Lambda$ in three dimensions, we have that
\begin{equation}
\eta^{npq}w_{pq} = 0.
\end{equation}
In consequence, the antisymmetric component of $w_{mn}$ is vanishing on shell. Then, the trace of $w_{mn}$, according to (\ref{21}) disappears on shell and also $w_{mn}$ has a null covariant divergence ($\nabla_m w^{mn} = 0$). Therefore, on shell, $w_{mn}$ is symmetric, transverse and traceless, which we denote by $h_{mn}^{Tt}$. These constraints reduce the nine components of $w_{mn}$ to two and the field equation boils down to
\begin{equation}
\eta^{mpq} \nabla_p h_q ^{nTt} - \mu h^{mnTt} = 0
\end{equation}
or
\begin{equation}
P_{mn}^{pq-}h_{pq}^{Tt} = 0 ,
\end{equation}
where
\begin{equation}
P_{mn}^{pq\pm} \equiv \frac{1}{4} [ ( \delta_m ^p \delta_n ^q + \delta_m ^q \delta_n ^p ) \pm (\eta_m ^{pr}\frac{\nabla_r}{\mu}\delta_n ^q +  \eta_m ^{qr}\frac{\nabla_r}{\mu}\delta_n ^p ) ],
\end{equation}
are parity-sensitive operators which allow the decomposition of the two components of $h_{mn}^{Tt}$:
\begin{equation}
h_{mn}^{Tt} = P_{mn}^{pq+}h_{pq}^{Tt} + P_{mn}^{pq-}h_{pq}^{Tt} ,
\end{equation}
in a such way that
\begin{equation}
h_{mn} = h_{mn}^{Tt+} + h_{mn}^{Tt-} \quad and \quad h_{mn}^{Tt+} - h_{mn}^{Tt-} = \frac{1}{\mu}(\eta_m ^{pq}\nabla_p h_{qn} + \eta_n ^{pq}\nabla_p h_{qm}) 
\end{equation}
In this way, the self-dual massive spin-2 keeps the propagation of one degree of freedom when is minimally coupled on (A)dS background, namely $h_{mn}^{Tt+}$, since $h_{mn}^{Tt-} = 0$ on shell. By iterating the field equation, we obtain that $h_{mn}^{Tt+}$ satisfies
\begin{equation}
[\Box - (\mu^2 + 3 \Lambda)] h_{mn}^{Tt+}= 0 .
\end{equation}

Having checked out that the self-dual massive gravity keeps the propagation of only one degree of freedom, we proceed to see the relationship with the partially masslees model in three dimensions. We rewrite the action (\ref{19}) as
\begin{equation}
I = \int d^3 x \sqrt{-g}[w_{mq} \eta^{mnp}\nabla_n \tilde{e}_p ^q - \frac{1}{2}\tilde{e}_{mq} \eta^{mnp}\nabla_n \tilde{e}_p ^q - \frac{1}{2}\sqrt{-g}\mu g^{mq}g^{np}(w_{mn}w_{pq} - w_{mq}w_{np})],
\end{equation}
where we have introduced an auxiliary field $\tilde{e}_{mn}$ such that after integrating out through its field equation:
$\eta^{mnp}\nabla_n \tilde{e}_p ^q = \eta^{mnp}\nabla_n w_p ^q $, we recover (\ref{19}). This is another way to rewrite the self-dual action for massive spin-2 field on (A)dS. In a flat spacetime , this action was introduced in ref.\cite{AK} from which a second-order action can be reached (after using the solution of the field equation derived from independent variations on $w_{mn}$) and this second-order action can be used to establish the equivalence with the third order topologically massive gravity \cite{ArKhSt}. As we will see, this equivalence will be broken on (A)dS because the topologically massive gravity does not enjoy the partially massless symmetry. Making the following redeﬁnitions
\begin{equation}
w_{mn} \rightarrow \frac{1}{\sqrt{\mu}}\omega_{mn} , \quad \tilde{e}_{mn} \rightarrow \sqrt{\mu}e_{mn},
\end{equation}
we obtain
\begin{equation}\label{33}
I = \int d^3 x [\omega_{mq} \epsilon^{mnp}\nabla_n e_p ^q - \frac{1}{2}\sqrt{-g}(\omega_{mn}\omega^{nm} - \omega^2)] - \frac{1}{2}\mu e_{mq} \epsilon^{mnp}\nabla_n e_p ^q ],
\end{equation}
which is just the partially massless three-dimensional action augmented by a Chern-Simons term for $e_{mn}$. This action is invariant under the partially massless gauge transformations (\ref{sym}). In this way, the self-dual massive gravity on (A)dS background is related to the partially massless theory in three dimensions. From (\ref{33}), we have that $w_{mn} \equiv W_{mn(e)}$ as given by (\ref{4}) is solution of the equation coming after realizing the independent variations on $w_{mn}$ and plugging into the action (\ref{33}), we obtain the following second-order action
\begin{equation}\label{34}
I = \int d^3 x [ \frac{1}{2}e_{mq} \epsilon^{mnp}\nabla_n W_{p(e)} ^q  - \frac{1}{2}\mu e_{mq} \epsilon^{mnp}\nabla_n e_p ^q ] .
\end{equation}
This action is clearly invariant under the partially massless symmetry and dual equivalent to the self-dual massive spin-2 on (A)dS. If we decompose $e_{mn}$ in its symmetric and antisymmetric parts ($e_{mn} = h_{mn} + \eta_{mnp}v^p $), it is straightforward to show that
\begin{equation}\label{35}
\eta^{mpq}\nabla_p W_{q(e)} ^n = - 2 G_{\text{lin}(h)}^{mn} - \Lambda (h^{mn} - g^{mn}h) + \Lambda \eta^{mnp}v_p ,
\end{equation}
where
\begin{equation}
G_{mn}^\text{lin} \equiv R_{mn}^\text{lin} - \frac{1}{2}g_{mn} R^\text{lin} - 2 \Lambda g_{mn} h
\end{equation}
is the linearized Einstein tensor with
\begin{equation}
R_{mn}^\text{lin} = -\frac{1}{2}[\nabla^2 h_{mn} + \nabla_m \nabla_n h - \nabla^p \nabla_m h_{np} - \nabla^p \nabla_n h_{mp}]
\end{equation}
and
\begin{equation}
R^\text{lin} = - [\nabla^2 h - \nabla_p \nabla_q h^{pq}] - 2 \Lambda h ,
\end{equation}
being the linearized Ricci and scalar curvature, respectively. Thus, the action (\ref{34}) is rewritten as
\begin{eqnarray}\label{42}
I = \int d^3 x \sqrt{-g}[ &-&\frac{1}{2}\nabla_p h_{mn} \nabla^p h^{mn} + \frac{1}{2} \nabla_p h \nabla^p h - \nabla_m h \nabla_n h^{mn} + \nabla_n h_{mp} \nabla^p h^{mn} {\nonumber} \\
&+& \frac{3}{2}\Lambda h_{mn}h^{mn} - \frac{1}{2}\Lambda h^2  - \frac{1}{2}\mu h_{mq} \eta^{mnp}\nabla_n h_p ^q + \mu v^m (\nabla^n h_{mn} - \nabla_ m h) {\nonumber} \\
&+&  \frac{1}{2}\mu v_m \eta^{mnp}\nabla_n v_p  - \Lambda v_{m}v^{m} ].
\end{eqnarray}

\section{New Model with Partially Massless}

In ref.\cite{AAK} the physical system, in flat spacetime, described  by the following action
\begin{equation}\label{43}
I = \int d^3 x [\frac{1}{2}e_{mq} \epsilon^{mnp}\partial_n W_{p(e)} ^q  - \frac{1}{2}\mu e_{mq} \epsilon^{mnp}\partial_n e_p ^q - \frac{1}{2}M^2 (e_{mn}e^{nm} - e^2)].
\end{equation}
was shown to describe the propagation of two healthy physical degrees of freedom with energy bounded from below and different masses t: $M^{\pm} = \pm\frac{1}{2}\mu + \sqrt{\frac{1}{4}\mu^2 + M^2}$. Also, this system was shown to be dual equivalent (\cite{ArKhSt}) to the dubbed General Massive Gravity (\cite{BHT}, \cite{BHT2}). In terms of the symmetric $(h_{mn})$ and antisymmetric $(v_m ) $ of $e_{mn}$, the action (\ref{43}) is written as
\begin{eqnarray}\label{44}
I = \int d^3 x [ &-&\frac{1}{2}\partial_p h_{mn} \partial^p h^{mn} + \frac{1}{2} \partial_p h \partial_p h - \partial_m h \partial_n h^{mn} + \partial_n h_{mp} \partial^p h^{mn} {\nonumber} \\
&-& \frac{1}{2}M^2 (h_{mn}h^{mn} -  h^2)  - \frac{1}{2}\mu h_{mq} \epsilon^{mnp}\partial_n h_p ^q + \mu v^m (\partial^n h_{mn} - \partial_ m h) {\nonumber} \\
&+&  \frac{1}{2}\mu v_m \epsilon^{mnp}\partial_n v_p  - M^2 v_{m}v^{m} ].
\end{eqnarray}

In this section, we will show that the generalization of (\ref{44}) on (A)dS enjoys of the partially massless symmetry and coincides with (\ref{42}), at this point. The action (\ref{43}) is written, on (A)dS background, as
\begin{eqnarray}\label{45}
I = \int d^3 x \sqrt{-g}[ &-&\frac{1}{2}\nabla_p h_{mn} \nabla^p h^{mn} + \frac{1}{2} \nabla_p h \nabla_p h - \nabla_m h \nabla_n h^{mn} + \nabla_n h_{mp} \nabla^p h^{mn} {\nonumber} \\
&-& \frac{1}{2}M^2 (h_{mn}h^{mn} -  h^2)  +2\Lambda (h_{mn}h^{mn} -  \frac{1}{2}h^2) - \frac{1}{2}\mu h_{mq} \eta^{mnp}\nabla_n h_p ^q +  {\nonumber} \\
&+&  \frac{1}{2}\mu v_m \eta^{mnp}\partial_n v_p + \mu v^m (\nabla^n h_{mn} - \nabla_ m h) - M^2 v_{m}v^{m} ].
\end{eqnarray}

 Now, we will show that this action describes the propagation of two degrees of freedom and how the partially massless symmetry arises from this action at the point $M^2 = \Lambda$ . The field equations which are obtained after independent variation on $h_{mn}$and $v_m$ are
\begin{eqnarray}\label{46}
\mathbb{G}_{mn} &\equiv& \Box h_{mn} + \nabla_m \nabla_n h - \nabla_p \nabla_m h_n ^p - \nabla_p \nabla_n h_m ^p - g_{mn}(\Box h- \nabla_p \nabla_q h^{pq}) {\nonumber} \\
&-& M^2 (h_{mn} - g_{mn}h) + 4\Lambda ( h_{mn} - \frac{1}{2}g_{mn}h) - \frac{\mu}{2}(\eta_m ^{pq}\nabla_p h_{qn} + \eta_n ^{pq}\nabla_p h_{qm}) {\nonumber} \\
&-& \frac{\mu}{2}(\nabla_m v_n + \nabla_n v_m ) + \mu g_{mn}\nabla_p v^p = 0
\end{eqnarray}
and
\begin{equation}\label{47}
F^m \equiv \eta^{mnp}\nabla_n v_p + (\nabla_n h^{mn} - \nabla^m h) -2\frac{M^2}{\mu}v^m = 0 .
\end{equation}
The trace of (\ref{46}) tell us that
\begin{equation}\label{48}
\mathbb{G} = (\Box h- \nabla_p \nabla_q h^{pq}) + 2(\Lambda - M^2)h - 2\mu \nabla_p v^p = 0,
\end{equation}
while the double covariant divergence of (\ref{46}) lead to
\begin{equation}\label{49}
\nabla_m \nabla_n \mathbb{G}^{mn} = M^2 (\Box h- g_{mn}\nabla_p \nabla_q h^{pq}) = 0 .
\end{equation}

From the covariant divergence of (\ref{47}), we obtain that
\begin{equation}\label{50}
\nabla_m F^m =  (\Box h- g_{mn}\nabla_p \nabla_q h^{pq}) - 2\frac{M^2}{\mu}\nabla_m v_m = 0 .
\end{equation}
Thus, by taking into account (\ref{48}), (\ref{49}) and (\ref{50}), we arrive to
\begin{equation}
(\Lambda - M^2)h = 0 .
\end{equation}

If $\Lambda - M^2 \neq 0$, we have that $h_{mn}$ is traceless: $h = 0$, and transverse: $\nabla_n h^{mn} = 0$, moreover, the vector field $v_m$ is vanishing on shell, therefore the field equation (\ref{46}) is reduced to
\begin{equation}
( \Box h_{mn} - 2\Lambda h_{mn} - M^2 h_{mn} - \frac{\mu}{2}(\eta_m ^{pq}\nabla_p h_{qn} + \eta_n ^{pq}\nabla_p h_{qm}) ) = 0
\end{equation}
and considering the decomposition of $h_{mn}$ in its two sensitive parity components ($h_{mn}^+ + h_{mn}^-$), we have the propagation of two degrees of freedom characterized by field equations:
\begin{equation}
[ \Box \mp \mu^2  -( M^2 + 2\Lambda )]h_{mn}^\pm  = 0 ,
\end{equation}
with masses given by $M^{\pm} = \pm\frac{1}{2}\mu + \sqrt{\frac{1}{4}\mu^2 + M^2}$.

On the other hand, the partially massless symmetry arises when: $\Lambda = M^2$. In fact, at this point, the action (\ref{45}) is just our second-order action for the self-dual massive gravity(\ref{42}), obtained in Sec. 5 with explicit partially massless symmetry.
We can modify action (\ref{34}), by introducing a new auxiliary field $v_{mn}$, as was realized in ref.\cite{ArKhSt} at flat spacetime:
\begin{equation}
I = \int d^3 x \frac{1}{2}e_{mq} \epsilon^{mnp}\nabla_n W_{p(e)} ^q  - \mu\int d^3 x v_{mq} \epsilon^{mnp}\nabla_n e_p ^q + \frac{1}{2}\mu\int d^3 x v_{mq} \epsilon^{mnp}\nabla_n v_p ^q .
\end{equation}

In fact, the elimination of the auxiliary field $v_{mn}$ by using its field equation:
\begin{equation}
\epsilon^{mnp}\nabla_n e_p ^q = \epsilon^{mnp}\nabla_n v_p ^q ,
\end{equation}
leads to recover the action (\ref{34}). On the other hand, independent variation on $e_{mn}$ yields
\begin{equation}
\epsilon^{mnp}\nabla_n W_{p(e)} ^q = \mu\epsilon^{mnp}\nabla_n v_p ^q
\end{equation}
and using this result, we arrive to a third-order action:
\begin{equation}
I = \int d^3 x -\frac{1}{2}e_{mq} \epsilon^{mnp}\nabla_n W_{p(e)} ^q  + \frac{1}{2\mu}\int d^3 x W_{mq(e)} \epsilon^{mnp}\nabla_n W_{p(e)} ^q .
\end{equation}
or
\begin{eqnarray}\label{59}
I &=& \int d^3 x [ -\frac{1}{2}h_{mn}G_\text{lin}^{mn} + \frac{1}{2\mu}h_{mn}C^{mn} ] \\{\nonumber}\
&-& \Lambda \int d^3 x \sqrt{-g}[\frac{1}{2\mu}e_{mq} \eta^{mnp}\nabla_n e_p ^q - \frac{1}{2}\sqrt{-g}\mu g^{mq}g^{np}(e_{mn}e_{pq} - e_{mq}e_{np})] ,
\end{eqnarray}
being
\begin{equation}
C^{mn} \equiv \eta^{mpq}\nabla_p (R_q ^n - \frac{1}{4}\delta_q ^n - 2\Lambda h_q ^n ) ,
\end{equation}
the linearized Cotton tensor, which is symmetric, freely divergence and traceless. If the spacetime is flat, this action is the linearized topologically massive gravity. But this is not the case, we have the topologically massive gravity augmented with a self-dual action for $e_{mn}$. On a flat spacetime, each of the two terms can make unstable the topologically massive gravity, the hamiltonian is not bounded from below (\cite{AAK}) although some restrictions of mass ratio might restore unitarity (\cite{varios1}). It will deserve an additional study of the action (\ref{59}).

\section{Conclusions}
We have seen how the scalar excitation of the partially massless in three dimensions can be understood as emerging from the linearized Einstein-Hilbert action, when it is turned on (A)dS gravitational background. This phenomenon was extended to higher dimensions, where mixed symmetry fields are locally trivial, for instance the Curtright field in four dimensions. Moreover, the self-dual massive gravity on (A)dS background looks like the partially massless theory coupled with a triadic Chern-Simons term, generates a change from the partially massless mode to a massive spin-2 mode. From the self-dual action, we have obtained a new physical model described by second-order field equations and having explicit partially massless symmetry.  Finally, we have seen that the equivalence with the purely topologically massive gravity on (A)dS is spoiled by a triadic self-dual action, as is indicated in (\ref{59}).

\section{ACKNOWLEDGMENTS}
We would like to Pio J. Arias for useful discussions.

\end{document}